\documentclass[prb,twocolumn,superscriptaddress,showpacs]{revtex4-1} % Change this later

\usepackage{amsmath} % Math package
\usepackage{amssymb} % Math symbols
\usepackage{graphicx} % Floats package
\usepackage{xcolor}	% Text color
\usepackage[lofdepth,lotdepth]{subfig} % Subfigures
\usepackage{multirow} % Allows multirow in tables
\usepackage[hyperindex,breaklinks]{hyperref} % URL package
\usepackage{placeins} % Prevent floats from leaking
\usepackage{standalone} % Inputs

\newcommand{\subfigimga}[3][,]{%
  \setbox1=\hbox{\includegraphics[#1]{#3}} % Store image in box
  \leavevmode\rlap{\usebox1} % Print image
  \rlap{\hspace*{-6pt}\raisebox{\dimexpr\ht1-2\baselineskip}{#2}} % Print label
  \phantom{\usebox1} % Insert appropriate spcing
}

\newcommand{\subfigimgb}[3][,]{%
  \setbox1=\hbox{\includegraphics[#1]{#3}} % Store image in box
  \leavevmode\rlap{\usebox1} % Print image
  \rlap{\hspace*{-2pt}\raisebox{\dimexpr\ht1-2\baselineskip}{#2}} % Print label
  \phantom{\usebox1} % Insert appropriate spcing
}

\begin{document}

\title{Critical analysis of the response function in low dimensional materials}
\author{Simon Divilov}
\email{simon.divilov@uam.es.}
\affiliation{Departamento de F\'{i}sica de la Materia Condensada and 
Condensed Matter Physics Center (IFIMAC).
Universidad Aut\'{o}noma de Madrid, Cantoblanco, 28049 Madrid.
Spain.}
\author{Sara G. Mayo}
\affiliation{Departamento de F\'{i}sica de la Materia Condensada and 
Condensed Matter Physics Center (IFIMAC).
Universidad Aut\'{o}noma de Madrid, Cantoblanco, 28049 Madrid.
Spain.}
\author{Jose M. Soler}
\affiliation{Departamento de F\'{i}sica de la Materia Condensada and 
Condensed Matter Physics Center (IFIMAC).
Universidad Aut\'{o}noma de Madrid, Cantoblanco, 28049 Madrid.
Spain.}
\author{Felix Yndurain}
\email{felix.yndurain@uam.es.}
\affiliation{Departamento de F\'{i}sica de la Materia Condensada and 
Condensed Matter Physics Center (IFIMAC).
Universidad Aut\'{o}noma de Madrid, Cantoblanco, 28049 Madrid.
Spain.}

\begin{abstract}
The presence of sharp peaks in the real part of the static dielectric response function are usually accepted as indication of charge or spin instabilities in a material. 
However, there are misconceptions that Fermi surface (FS) nesting guarantees a peak in the response function like in one-dimensional systems, and, in addition, response function matrix elements between empty and occupied states are usually considered of secondary importance and typically set to unity like in the free electron gas case.
In this work, we explicitly show, through model systems and real materials, within the framework of density functional theory, that predictions about the peaks in the response function, using FS nesting and constant matrix elements yields erroneous conclusions.
We find that the inclusion of the matrix elements completely alters the structure of the response function. In all the cases studied other than the one-dimensional case we find that the inclusion of matrix elements washes out the structure found with constant matrix elements.
Our conclusion is that it is imperative to calculate the full response function, with matrix elements, when making predictions about instabilities in novel materials.
\end{abstract}

\pacs{71.15.Dx, 71.15Mb, 71.18+y, 71.30+h, 71.45.Lr}

\maketitle
	\section{Introduction}
    	Electronic correlation in 1D and 2D low dimensional solids  has been a highly active area of research for some time now.\cite{10.1038/nphys253}
In particular, for  high \textit{T}\textsubscript{c} superconductivity to exist, the material must exhibit electronic, magnetic or ionic instabilities which can also manifest as charge density waves, spin density waves or lattice distortions, respectively. 
The instabilities lead to a positive attractive potential between the electrons, allowing for the formation of bounding pairs either in real or k-space.
In addition, these instabilities are driven by coherence effects where there is a large concentration of electrons at the same energy value.
For this reason, large peaks in the density of states (DOS) near the Fermi energy can be associated with possible instabilities that can lead to either superconductivity or density waves (either spin or charge) or mixture of them.
Low dimensional materials have an added benefit that their DOS diverges at a Van Hove singularity\cite{PhysRev.89.1189} (VHS) at saddle points making them prime candidates for high \textit{T}\textsubscript{c} superconductivity and other broken symmetries. In addition, in low dimensional systems parallel portions of the Fermi surface (line) indicate a quasi one-dimensional behavior and therefore a tendency to electronic instability.
Hence, in the literature, the Fermi surface (FS) nesting, or, regions of the Fermi surface that are parallel to each other for a specific \textbf{q} vector, are used to determine whether an instability exists.\cite{WHANGBO96,PhysRevLett.86.5100}

A common misconception is that a peak in the DOS guarantees an instability, such as a charge density wave.\cite{MARKIEWICZ19971179}
However, the actual strength of an instability is determined by a sharp peak in the real part of the static electric response function or susceptibility $\chi(\textbf{q})$.
In fact, M. D. Johannes \textit{et al}.\cite{PhysRevB.77.165135} have shown that using the FS nesting as an indicator for instabilities will often lead to misleading and incorrect results.
Then clearly, it is imperative to properly evaluate the response function for making predictions about %superconductivity
correlated states in the materials.
Unfortunately, evaluation of $\chi(\textbf{q})$, both experimentally and computationally, has proven to be a difficult task.\cite{PhysRevB.77.165135}

In this work we attempt to elucidate the calculation of the response function.
We begin by discussing computational difficulties and pitfalls associated with calculating the response function and how to resolve them.
Then, using the techniques we developed, we evaluate the response function for simple models and real materials, all within the framework of density functional theory (DFT).
Finally we discuss promising applications of our technique to controversial cases in the literature. 

    \section{Response function}
	    In linear response theory a material is said to be unstable if the relation
\begin{equation}
V_{C}(\textbf{q})\chi(\textbf{q},\omega=0) \leq 0
\end{equation}
is violated.\cite{keldysk1990}
Here $\chi(\textbf{q},\omega)$ is the response function and $V_{C}(\textbf{q})=\frac{4\pi e^{2}}{|\textbf{q}|^{2}}$ is the Coulomb potential in momentum space.
In general, the response function is difficult to calculate, so the random phase approximation (RPA) is generally accepted, that assumes the electrons only feel the total electrostatic potential.
Then the response function may be written as\cite{gorkov1989}
\begin{equation}
\chi(\textbf{q},\omega) = \frac{\chi_0(\textbf{q},\omega)}{1-V_C(\textbf{q})\chi_0(\textbf{q},\omega)}
\end{equation}
where $\chi_0$ is the non-interacting response function.
For a general material $\chi_0$ can be written as\cite{PhysRev.115.786,PhysRev.109.741}
\begin{equation}
\chi_0(\textbf{q},\omega) = \sum_{ll'}\sum^{BZ}_{\textbf{k}}\frac{f(\epsilon_{l\textbf{k}},\mu) - f(\epsilon_{l'\textbf{k}+\textbf{q}},\mu)}{\epsilon_{l\textbf{k}}-\epsilon_{l'\textbf{k}+\textbf{q}} - \omega - i\delta} \left | F_{ll'\textbf{k}\textbf{k}+\textbf{q}} \right |^2
\label{eq:chi0}
\end{equation}
and the matrix elements
\begin{equation}
F_{ll'\textbf{k}\textbf{k}+\textbf{q}} = \langle l,\textbf{k} |e^{i\textbf{q}\cdot\textbf{r}}| l',\textbf{k}+\textbf{q} \rangle
\label{eq:overlap}
\end{equation}
where $f$ is the Fermi distribution function, $\mu$ is the chemical potential, $\epsilon_{l,\textbf{k}}$ is the single particle Kohn-Sham energy and $\langle\textbf{r}|l,\textbf{k}\rangle=u_{l\textbf{k}}(\textbf{r})e^{i\textbf{k}\cdot\textbf{r}}$ is the Bloch wave function.
Here $\mu$ is a free parameter, defined relative to the Fermi energy, to simulate a charged system.
It is straightforward to show that a system is unstable if the static non-interacting response function diverges and therefore we will focus on evaluating $\chi_0(\textbf{q},\omega=0)$, which for brevity, we will refer to as the response function.

A key element to properly evaluate the response function is to perform the calculation in the extended zone (non-periodic) scheme, rather than the periodic zone scheme, the typical convention in first principle codes.
This means that the response function is non-periodic due to the contribution of the matrix elements in $F$.
Following Ref. \citenum{10.1088/1361-648X/ab6e8e} one can write the \textbf{q} vector as a sum of \textbf{K}\textsubscript{\textbf{q}}, a unique vector inside the Brillioun zone (BZ) and \textbf{G}\textsubscript{\textbf{q}}, a unique reciprocal lattice vector.
Then all one needs to calculate the response function, for an arbitrary \textbf{q} vector, are the Fourier coefficients of the Bloch wave functions $\tilde{u}_{l\textbf{k}}(\textbf{G})$.
Using this method, Eq. \ref{eq:overlap} becomes
\begin{equation}
F_{ll'\textbf{k}\textbf{k}+\textbf{q}} =  \Omega\sum^{\textbf{G}_{\text{cut}}}_{\textbf{G}}\tilde{u}^{*}_{l\textbf{k}}(\textbf{G})\tilde{u}_{l'\textbf{k}+\textbf{K}_{\textbf{q}}}(\textbf{G}+\textbf{G}_{\textbf{q}})
\label{eq:overlapnew}
\end{equation}
where $\Omega$ is the simulation cell volume.
In this scheme, we have a number quantities that we must converge with.
(\textit{i}) The \textit{k}-grid of the BZ, (\textit{ii}) the energy cutoff for the Fourier expansion $|\textbf{G}_{\text{cut}}|^{2}$ and (\textit{iii}) the number of bands.

(\textit{i}) A good starting guess for convergence with the \textit{k}-grid is to use the same \textit{k}-grid that would give an accurate density of states (DOS) of the system.
The reasoning behind this is that a Taylor expansion of $\chi_0$ around $\textbf{q}=0$ yields the DOS evaluated at $\mu$.\cite{davidpines1989}
This implies that a converged DOS should yield a converged response function, with respect to the \textit{k}-grid.
A sign that the calculation is not converged can be deduced if $\chi_0(\textbf{q})$ in the constant matrix approximation ($F=1$), lacks the symmetry of the material in reciprocal space.
%
%For example, for the band structure of single layer graphene, the group of \textbf{k} vectors around the $\Gamma$ point, belong to $D_{6h}$,\cite{PhysRevB.85.115418} therefore one should expect, for the constant matrix response function, the group of \textbf{q} vectors, around the $\Gamma$ point, to belong to that same group.

(\textit{ii}) Much like for a plane-wave basis set, the converged cutoff energy must be determined by repeating the calculation for an increasing cutoff energy.
Since the only dependence of \textbf{G} is in the matrix elements, it is sufficient to calculate $F$ for a small number of bands and a coarse \textit{k}-grid to determine if you have a converged basis set.

(\textit{iii}) In principle the sum in Eq. \ref{eq:chi0} runs over all the bands to infinity, however in practice this would yield erroneous results.
This is because in DFT, continuum functionals poorly describe the highly excited states and break down completely for the Rydberg states.\cite{10.1063/1.477711} In addition, using localized orbital basis reproduces very poorly the high energy conduction bands.
Therefore, summing over an increasing number of conduction states will lead to an accumulation of errors in the response function.
For this reason a cutoff energy for the conduction bands $E^{c}_{\text{cut}}$ should be introduced to avoid summing over the high energy transitions.

Lastly, the response function is insensitive to the $\delta$ term, however it should be small enough to avoid artificially broadening the peaks.
In addition, while the $\delta$ term avoids a division by zero, it does not yield the correct answer at the $\Gamma$ points.
We found that a linear interpolation, using the surrounding points, is enough to correctly describe the response function there.

		\label{sec:respfunc}
	\section{Calculations}
	        We have performed first-principle calculations for model systems as well as for real materials using \textsc{siesta},\cite{Soler2002,10.1063/5.0005077} a numerical atomic orbital code.
The two model systems that we have studied are a hydrogen chain and a hydrogen square lattice.
The calculations were done with the local density approximation (LDA) using norm-conserving psuedopotentials generated by the Trouiller-Martins (TM)\cite{Troullier1991} scheme and a double-$\zeta$ basis set.
The BZ was sampled using a 300 and 90 000 \textit{k}-point Monkhorst-Pack (MP)\cite{Monkhorst1976} mesh for the hydrogen chain and hydrogen square lattice, respectively, with a mesh cutoff energy of 250 Ry.
For the response function, we have used 300 and 10 000 \textit{k}-points in the BZ for the H chain and H square lattice, respectively.

The real one-dimensional material that we have chosen to study was \textit{trans}-polyacetylene, while the two-dimensional materials were twisted bilayer graphene (tBLG) and monolayer VSe\textsubscript{2}.
The calculations were done with the generalized gradient approximation (GGA)\cite{PhysRevB.45.13244} including van der Waals interaction with the Berland and Hyldgaard functional,\cite{PhysRevB.89.035412} using norm-conserving psuedopotentials generated by the TM scheme and a double-$\zeta$ basis set.
The BZ was sampled using a 300, 900 and 40 000 \textit{k}-point MP mesh, with a mesh cutoff energy of 200 Ry for \textit{trans}-polyacetylene, tBLG and VSe\textsubscript{2}, respectively.
For the response function, we have used 300, 1 600 and 10 000 \textit{k}-points in the BZ for \textit{trans}-polyacetylene, tBLG and VSe\textsubscript{2}, respectively.
We found that the functionals and convergence parameters used for the real materials give results in good agreement with previous calculations.\cite{PhysRevB.99.045423, Feng2018, Coelho2019}
The response function parameters, unless otherwise stated, were $|\textbf{G}_{\text{cut}}|^{2}=$ 20 Ry, $E^{c}_{\text{cut}}=$ 15 eV, $\delta=$ 1$\times$10\textsuperscript{-11} eV and 0.025 eV for the temperature in the Fermi distribution function. All the valence band are included in the calculations.

			\subsection{Model systems}
\begin{figure} 	
    \centering
    \subfloat{\subfigimga[width=0.5\linewidth]{(\textbf{a})}{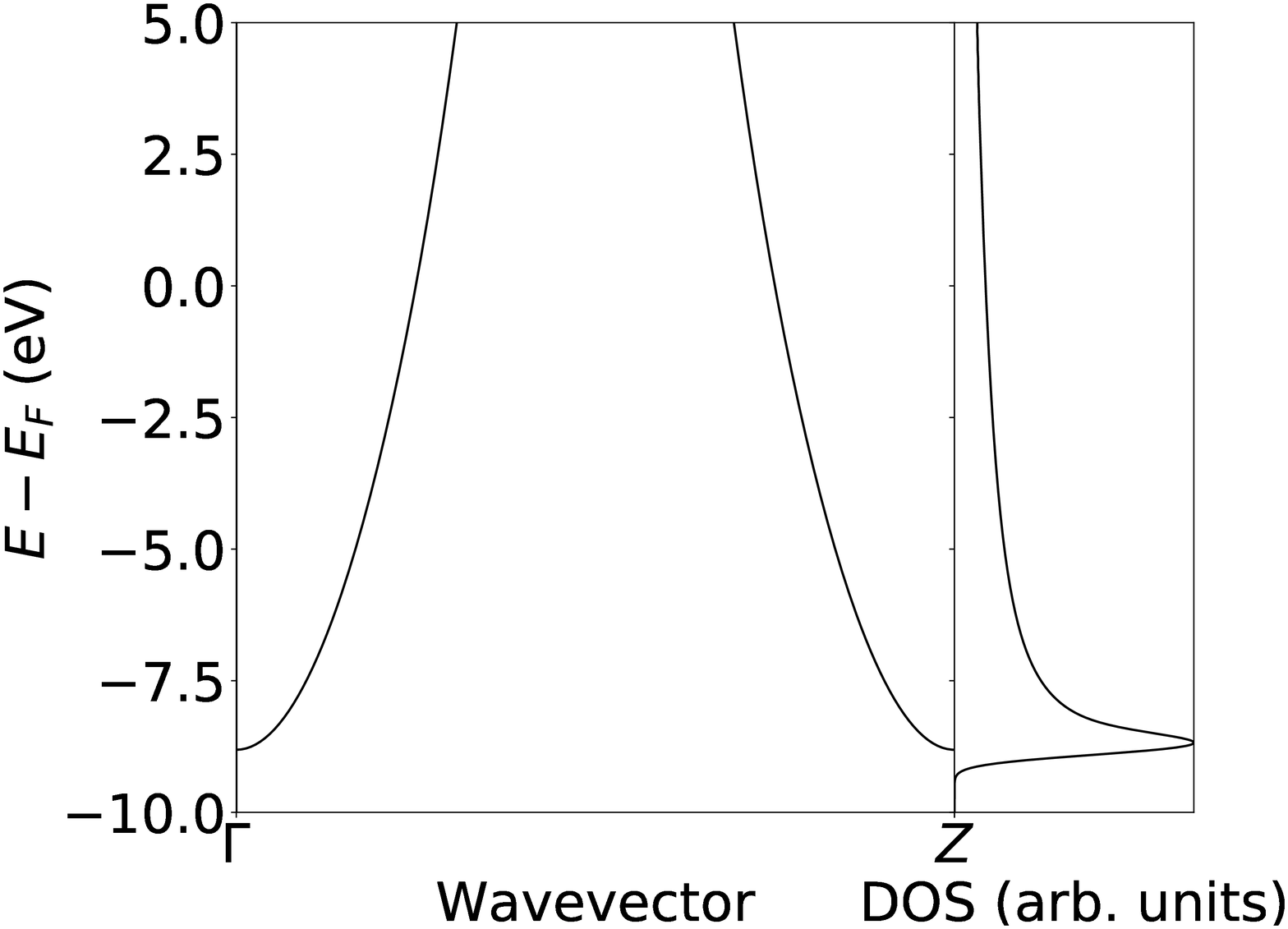}\label{fig:h_chain_bands}}%
    \vspace{-1.0\baselineskip}%
    \\
    \subfloat{\subfigimga[width=0.5\linewidth]{(\textbf{b})}{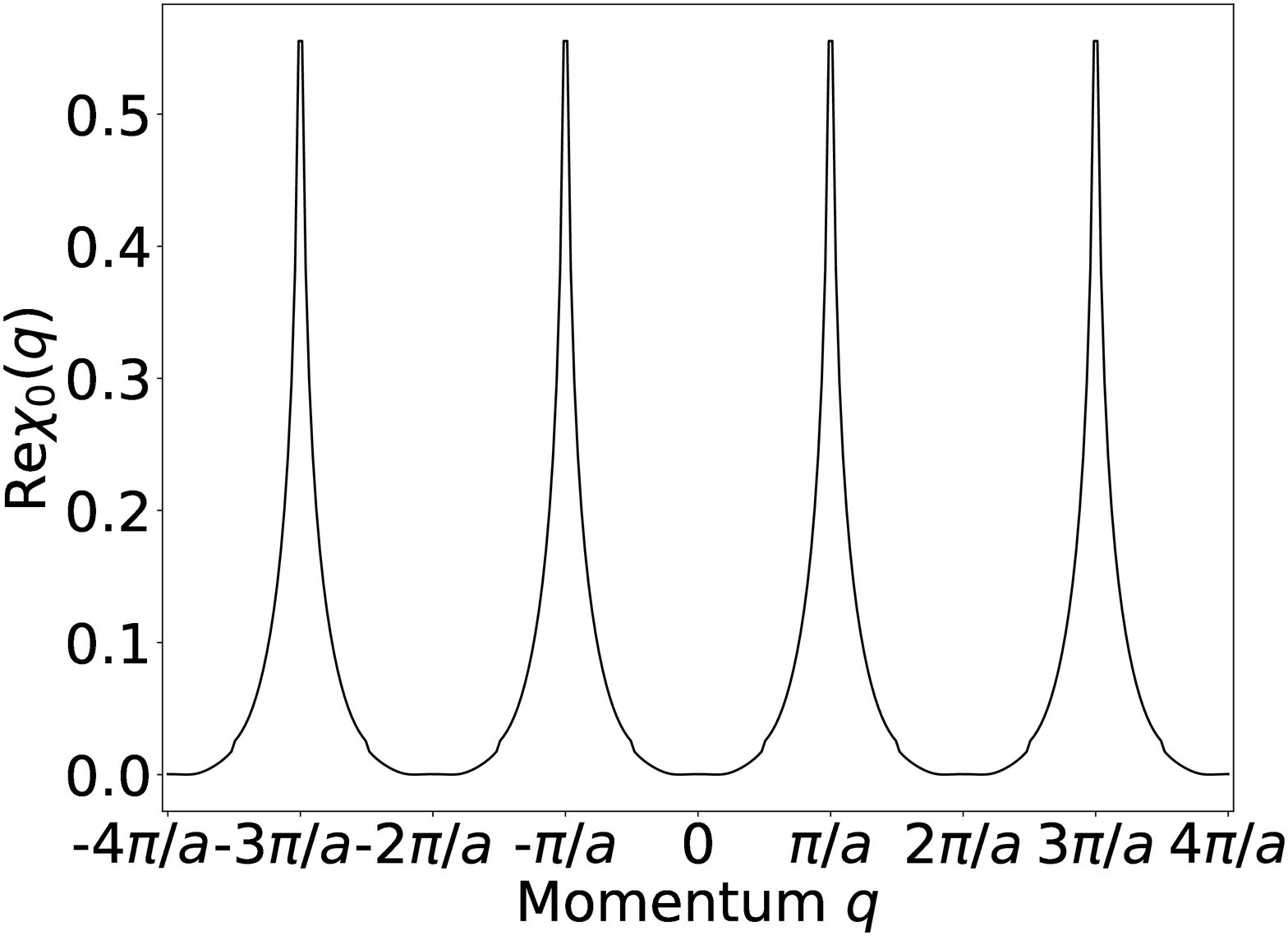}\label{fig:h_chain_cm}}%
    \vspace{-1.0\baselineskip}%
    \\    		
    \subfloat{\subfigimga[width=0.5\linewidth]{(\textbf{c})}{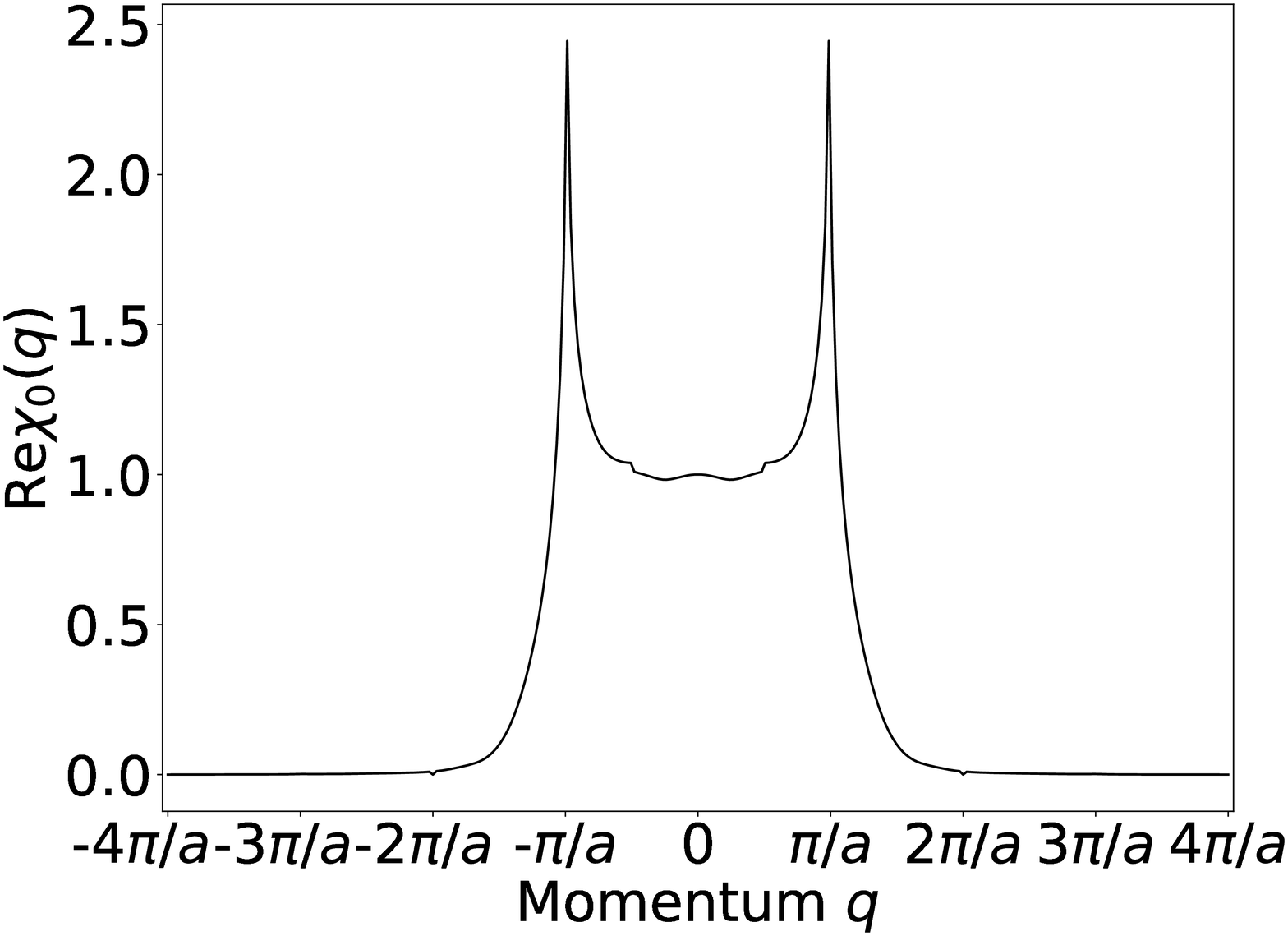}\label{fig:h_chain_chi}}%
    \caption{Response function of a 1D chain of hydrogen atoms. Panel (a) shows the band structure and the density of states. Panel (b) shows the real part of the response function assuming constant matrix elements, the lowest value is taken as the reference zero. In panel (c) the real part of the full response function scaled by $\operatorname{Re}\chi_0(\textbf{q}=0)$ is shown. In all cases the response function is evaluated at $\mu=0.0$ eV.}
    \label{fig:h_chain}
\end{figure}

			    The hydrogen chain and hydrogen square lattice are quintessential examples of simple low dimensional materials that most closely resemble tight-binding calculations.
Understanding how to process the results for these simple, yet non-trivial systems will provide us with a guide on how to handle real materials.
The hydrogen chain, at the calculated equilibrium lattice constant of 1.026 \AA{}, is described by an \textit{s}-band as shown in the band structure of Fig. \ref{fig:h_chain_bands}.
The real part of the response function for constant matrix elements ($F=1$), shown in Fig. \ref{fig:h_chain_cm}, is calculated relative to the minimum value within the BZ.
This is done so it is easier to identify the peaks in the response function, since we are searching for \textit{q}-vectors at which the material is unstable.
In addition, since the single particle eigenenergies are calculated in the periodic zone scheme, the constant matrix response function is periodic as well.
However, when matrix elements are included, as shown in Fig. \ref{fig:h_chain_chi}, the periodicity in \textit{q}-space is destroyed, but the position of the peaks within the first BZ are preserved.
The peaks at $q=\pm\pi/a$ signify the instability of the half filled \textit{s}-band because there is only one hydrogen atom per unit cell.
The instability leads to the well known dimerization, or Peierls distortion, of the hydrogen one dimensional lattice, where an energy band gap is created at multiples of $\pm\pi/2a$.
We also note that the full response function resembles the response function for a non-interacting electron gas\cite{1111.5337} due to the hydrogen chain parabolic-like character of the \textit{s}-band.
As we shall show, the preservation of the peaks when matrix elements are included is not a general feature that can be taken for granted.

%-----------------------------------------

\begin{figure}
    \centering
    \subfloat{\subfigimgb[width=0.5\linewidth]{(\textbf{a})}{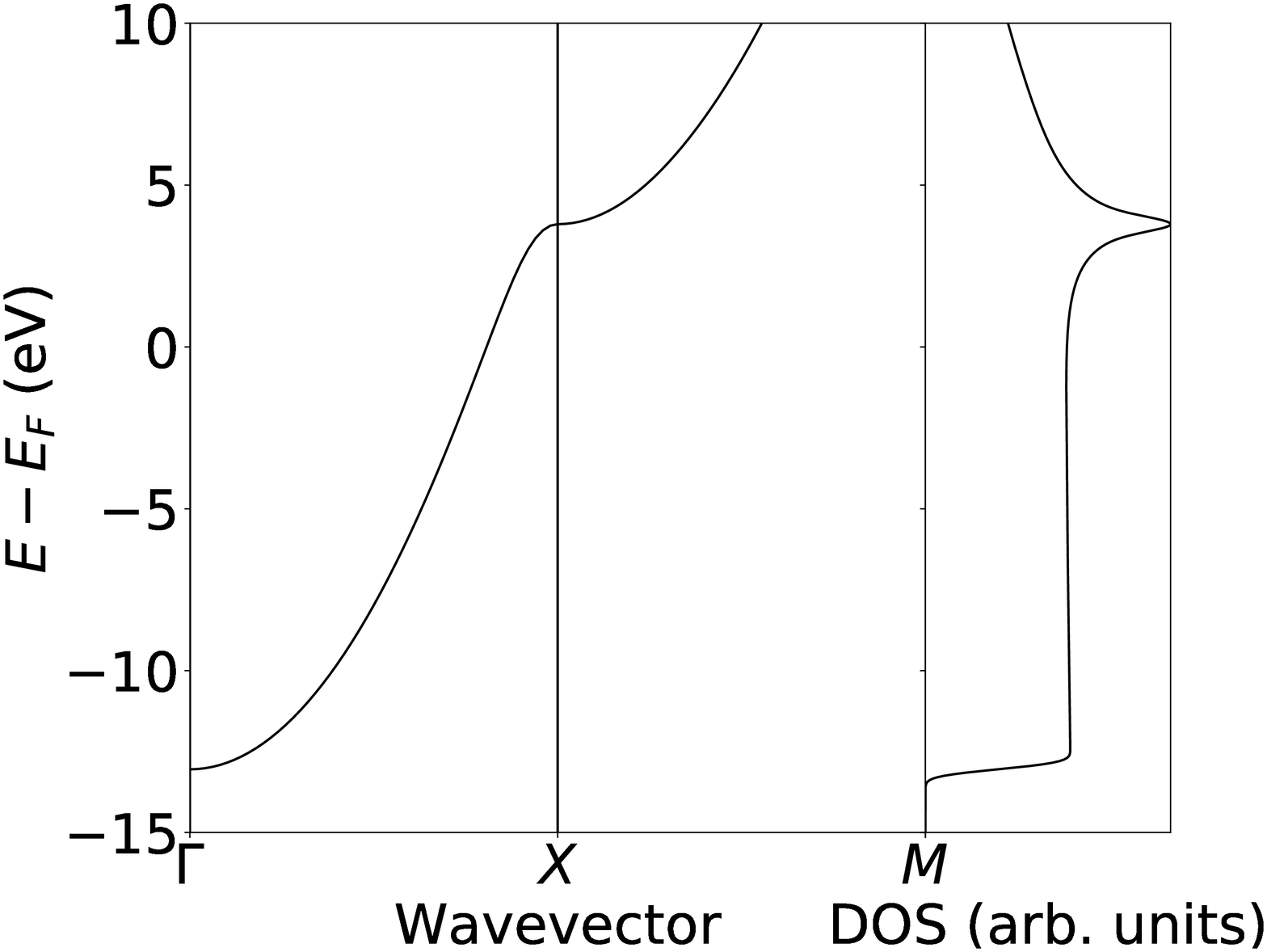}\label{fig:h_plane_bands}}%
    \hspace*{-15.0pt}%
    \vspace{-0.5\baselineskip}%
    \subfloat{\subfigimgb[width=0.5\linewidth]{(\textbf{b})}{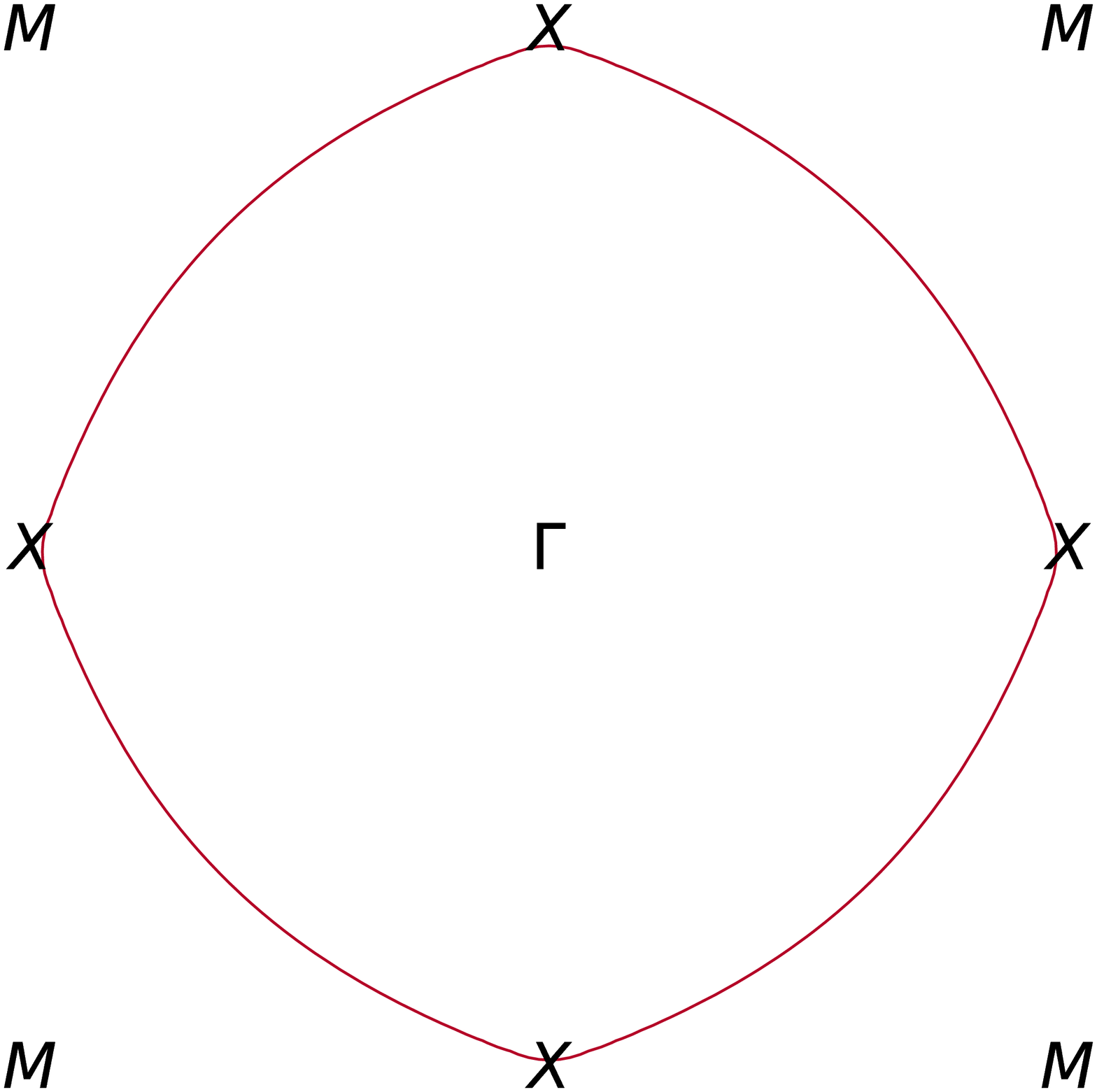}\label{fig:h_plane_fs}}%
    \vspace{-0.5\baselineskip}%
    \\	
    \subfloat{\subfigimgb[width=0.5\linewidth]{(\textbf{c})}{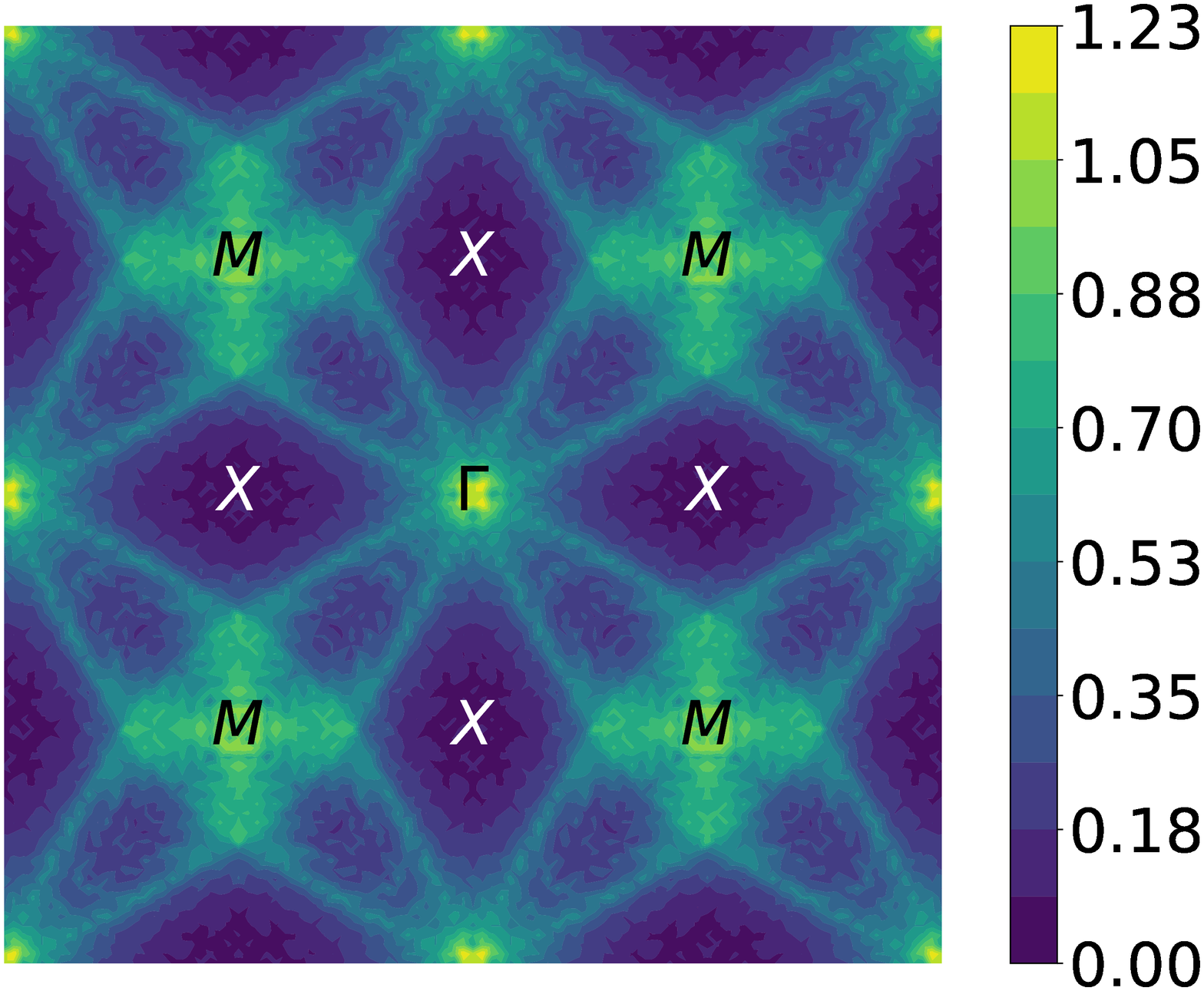}\label{fig:h_plane_cm}}%
    \hspace*{-15.0pt}%
    \subfloat{\subfigimgb[width=0.5\linewidth]{(\textbf{d})}{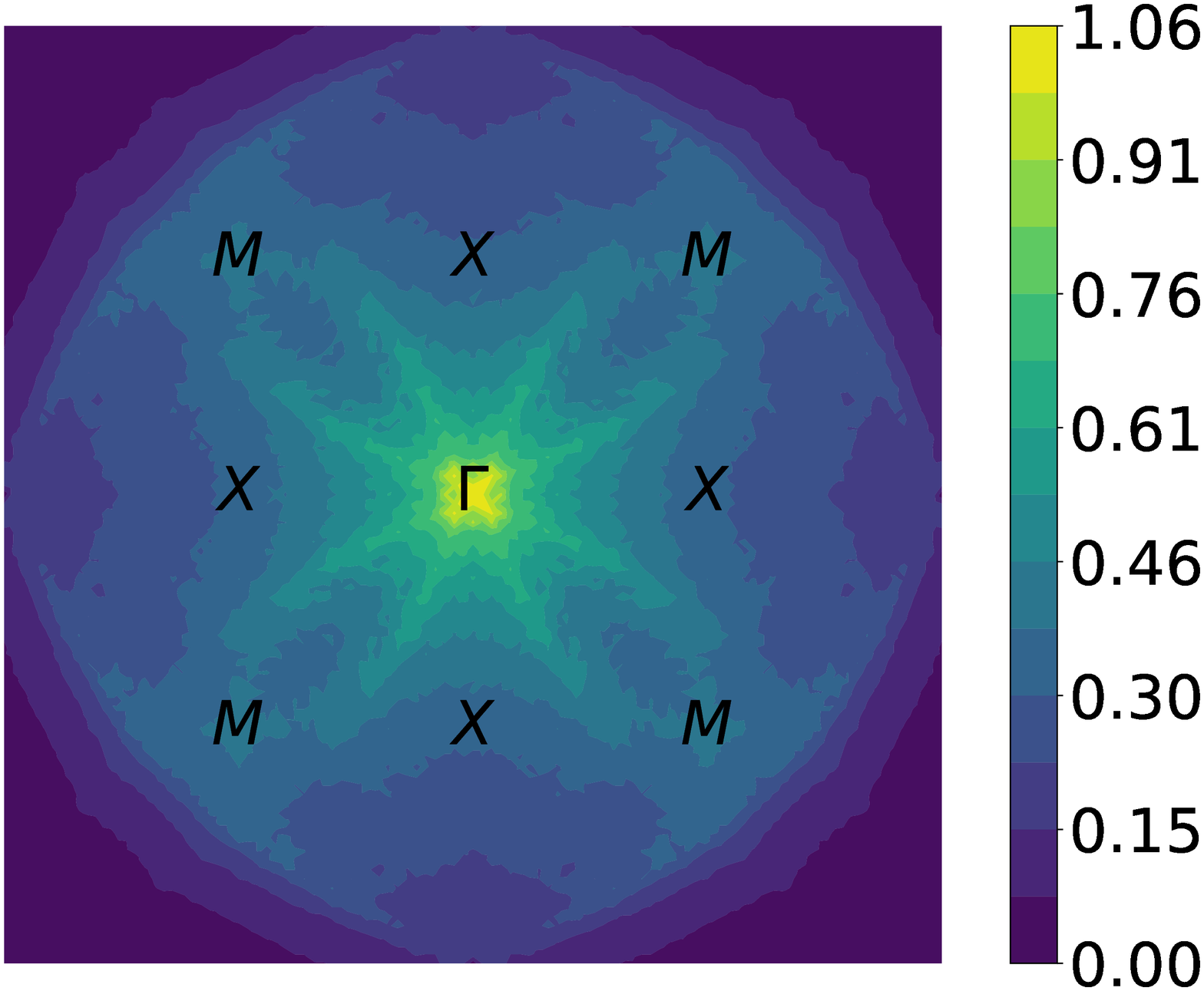}\label{fig:h_plane_chi}}%
    \caption{(Color on line) Response function of a hydrogen square lattice. Panel (a) shows the band structure and the density of states, the saddle point is at X point at 3.72 eV.  Panel (b) shows the Fermi surface when  $\mu=3.72$ eV.  Panel (c) is the real part of the constant matrix response function, where the lowest value was taken as zero, and in panel (d) the real part of the full response function scaled by $\operatorname{Re}\chi_0(\textbf{q}=0)$ is shown.}%The Fermi surface and response function were evaluated at $\mu=3.72$ eV.
    \label{fig:h_plane}
\end{figure}

%-----------------------------------------
We perform a similar set of calculations for the hydrogen square lattice as we have done for the hydrogen chain.
The hydrogen square lattice, at the calculated equilibrium lattice constant of 1.335 \AA{}, has a VHS due to the saddle point in the band structure at 3.72 eV, as shown in Fig. \ref{fig:h_plane_bands}.
The FS at the VHS, shown in Fig. \ref{fig:h_plane_fs}, exhibits nesting between the points $\{(\pm\pi/a,0),(0,\pm\pi/a)\}$ and the points $\{(\pm\pi/a,0),(0,\mp\pi/a)\}$, which can be equated to the $M$ points.
In addition, due to the saddle point, there is nesting at the $\Gamma$ point.
Likewise, in the constant matrix response function, shown in Fig. \ref{fig:h_plane_cm}, there is a clear peak at the $\Gamma$ point and $M$ points.
We note, as previously discussed in Ref. \citenum{PhysRevB.77.165135}, equating the FS nesting with peaks in the response function should be avoided, but for the case of the hydrogen square lattice this relation holds. 
However, when matrix elements are included, as shown in Fig. \ref{fig:h_plane_chi}, the peak at the $M$ points is completely washed out.
Evidently, the magnitude of the peak was not large enough since the matrix elements decay to zero away from the $\Gamma$ point and hence, unlike the hydrogen chain, there is no instability in the hydrogen square lattice at the VHS.
This remarkable result is in agreement with two-dimensional tight-binding calculations which found deviations from perfect FS nesting suppress the Peierls distortion.\cite{Yuan2001}
This case explicitly highlights that calculating the constant matrix element response function, even when in agreement with the FS nesting, cannot be used as a replacement for the full response function when discussing instabilities. 

%-----------------------------------------

			    \label{sec:model}
			\subsection{Real materials}
			    \subsubsection{Trans-polyacetylene}
\begin{figure} 	
    \centering
    \subfloat{\subfigimga[width=0.5\linewidth]{(\textbf{a})}{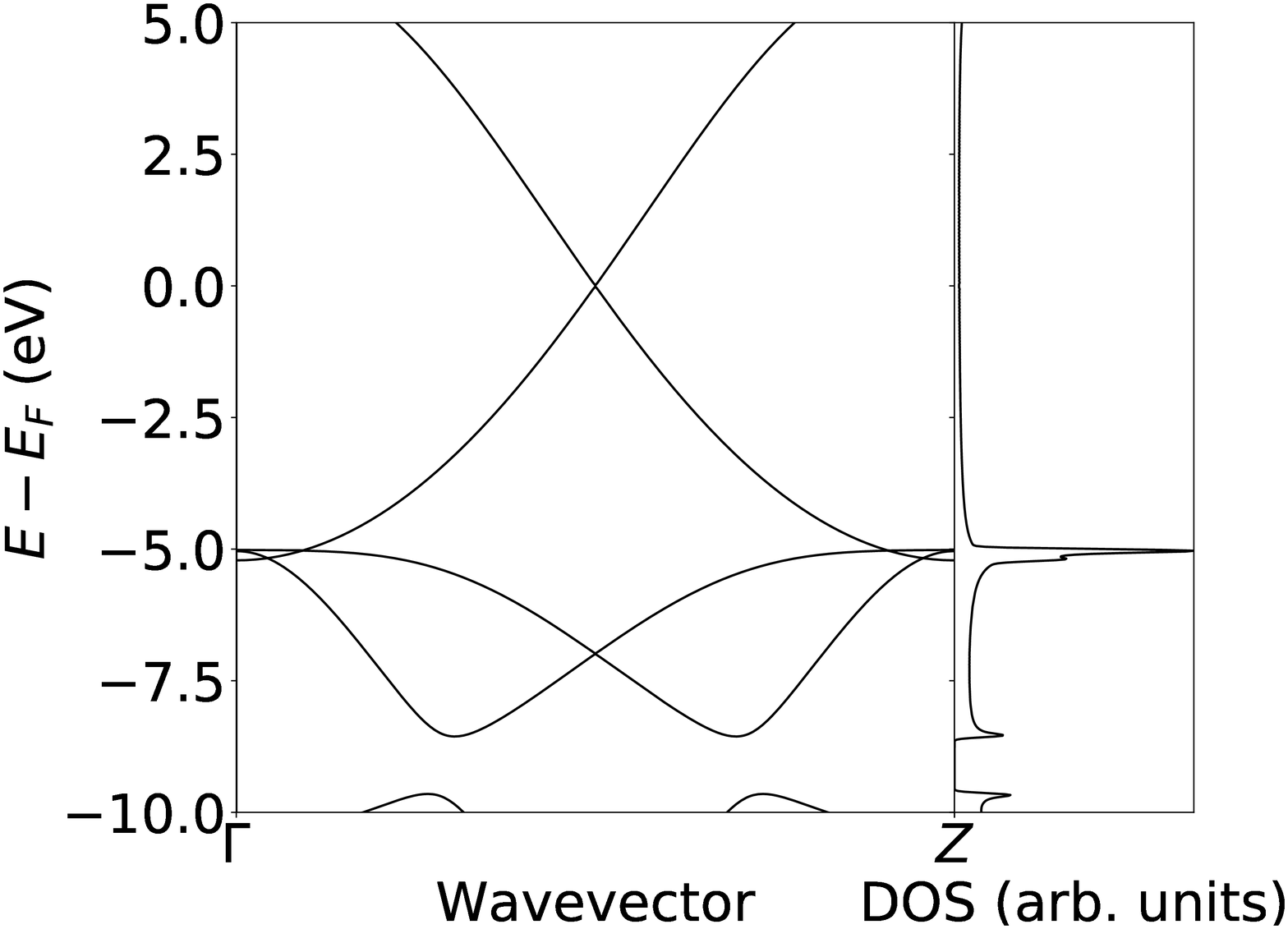}\label{fig:polyace_bands}}%
    \vspace{-1.0\baselineskip}%
    \\
    \subfloat{\subfigimga[width=0.5\linewidth]{(\textbf{b})}{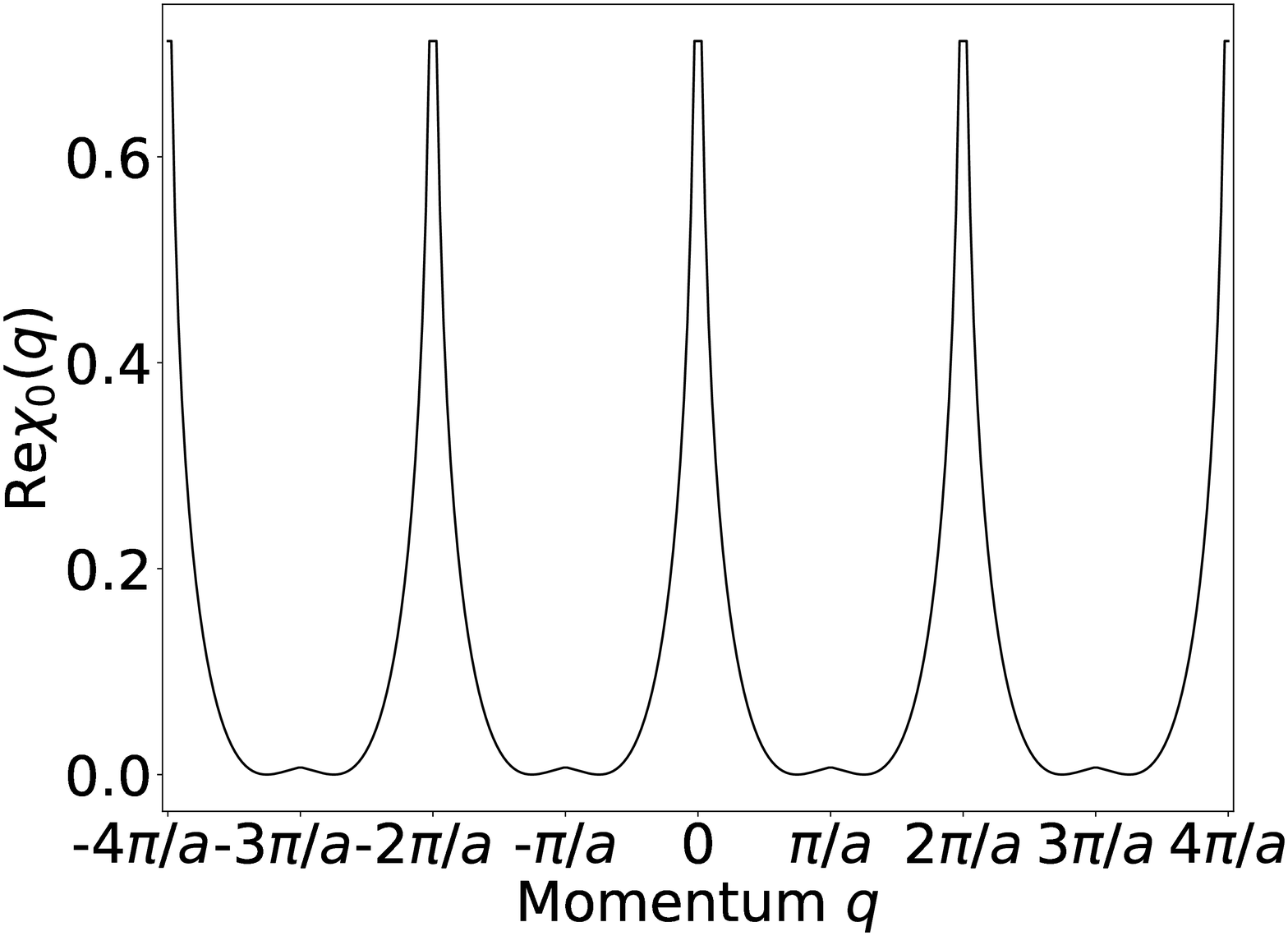}\label{fig:polyace_cm}}%
    \vspace{-1.0\baselineskip}%
    \\    		
    \subfloat{\subfigimga[width=0.5\linewidth]{(\textbf{c})}{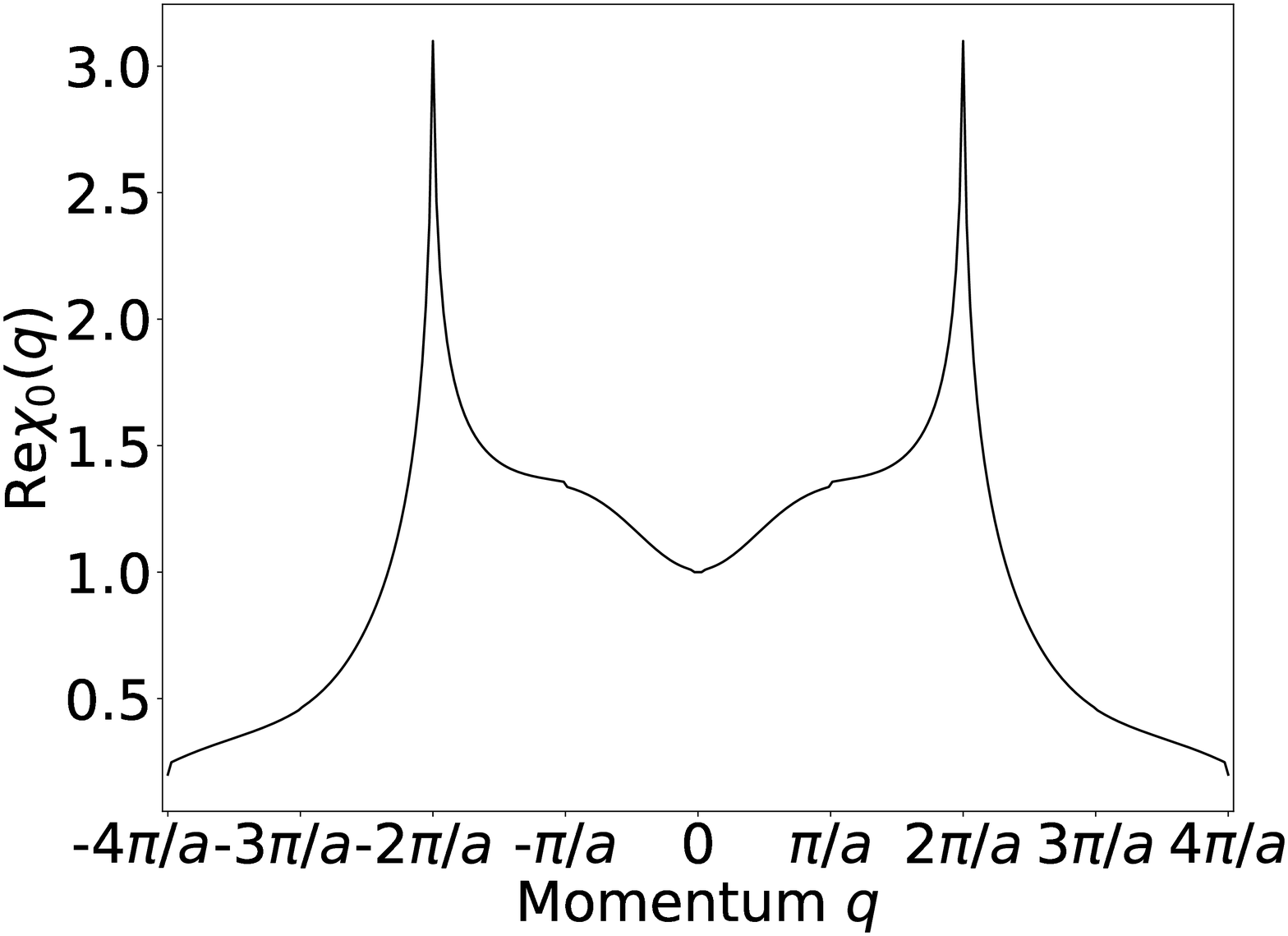}\label{fig:polyace_chi}}%
    \caption{Polyacetylene response function. Panel (a) shows the band structure. Panel (b) shows the real part of the response function assuming constant matrix elements, where the lowest value was taken as zero. In panel (c) the real part of the full response function scaled by $\operatorname{Re}\chi_0(\textbf{q}=0)$ is shown. The response function is evaluated at $\mu=0.0$ eV.}
    \label{fig:polyace}
\end{figure}

\textit{Trans}-polyacetylene is a one-dimension chain of C\textsubscript{2}H\textsubscript{2} where, dimerization in the C-C bond, has been observed\cite{LEFRANT1979191} and theoretically predicted to lead to the formation of neutral and charged solitons.\cite{PhysRevB.22.2099}
The calculation was performed for the undimerized chain at the relaxed atomic positions, calculated from first principles.
The calculated C-H and C-C bond lengths were 1.105 \AA{} and 1.402 \AA{}, respectively and the calculated C-C-C and C-C-H bond angles were 123.5$^{\circ}$ and 118.3$^{\circ}$, respectively.
Since the linear chain is undimerized, the system does not exhibit a band gap, as shown in Fig. \ref{fig:polyace_bands} and the linear dispersion near the Fermi energy is due to the $\pi$ electrons (p\textsubscript{z} orbitals).
In this system, we expect peaks in the response function because the dimerization has been observed experimentally.
Indeed, we found peaks in the constant matrix function, evaluated at the Fermi energy, shown in Fig. \ref{fig:polyace_cm}.
The large peaks at multiples of $q=\pm2\pi/a$ can be associated with the $\pi$ electrons, while the bumps at multiples of $q=\pm\pi/a$ are due to the $\sigma$ electrons.
When matrix elements are included, as shown in Fig \ref{fig:polyace_chi}, the peaks of the $\pi$ electrons is preserved, signifying that the linear chain is unstable and prefers to dimerize.
However the peaks in this system occur at $\pm2\pi/a$ rather than $\pm\pi/a$ as was the case for the hydrogen chain.
This is easily understood because in the primitive unit cell there are two CH groups, equivalent by a $\pi$ rotation, therefore, unlike the hydrogen chain, dimerization of the bonds can occur without modifying the unit cell.
In other words, the dimerization is a Peierls distortion at constant volume with the formation of band gaps at multiples of $\pm\pi/a$.
We note the dip near $q=0$ of the full response function is due to the inclusion of the $\sigma$ electrons.

%-----------------------------------------

%-----------------------------------------

\subsubsection{Twisted bilayer graphene}
\begin{figure} 	
    \centering
    \subfloat{\subfigimgb[width=0.5\linewidth]{(\textbf{a})}{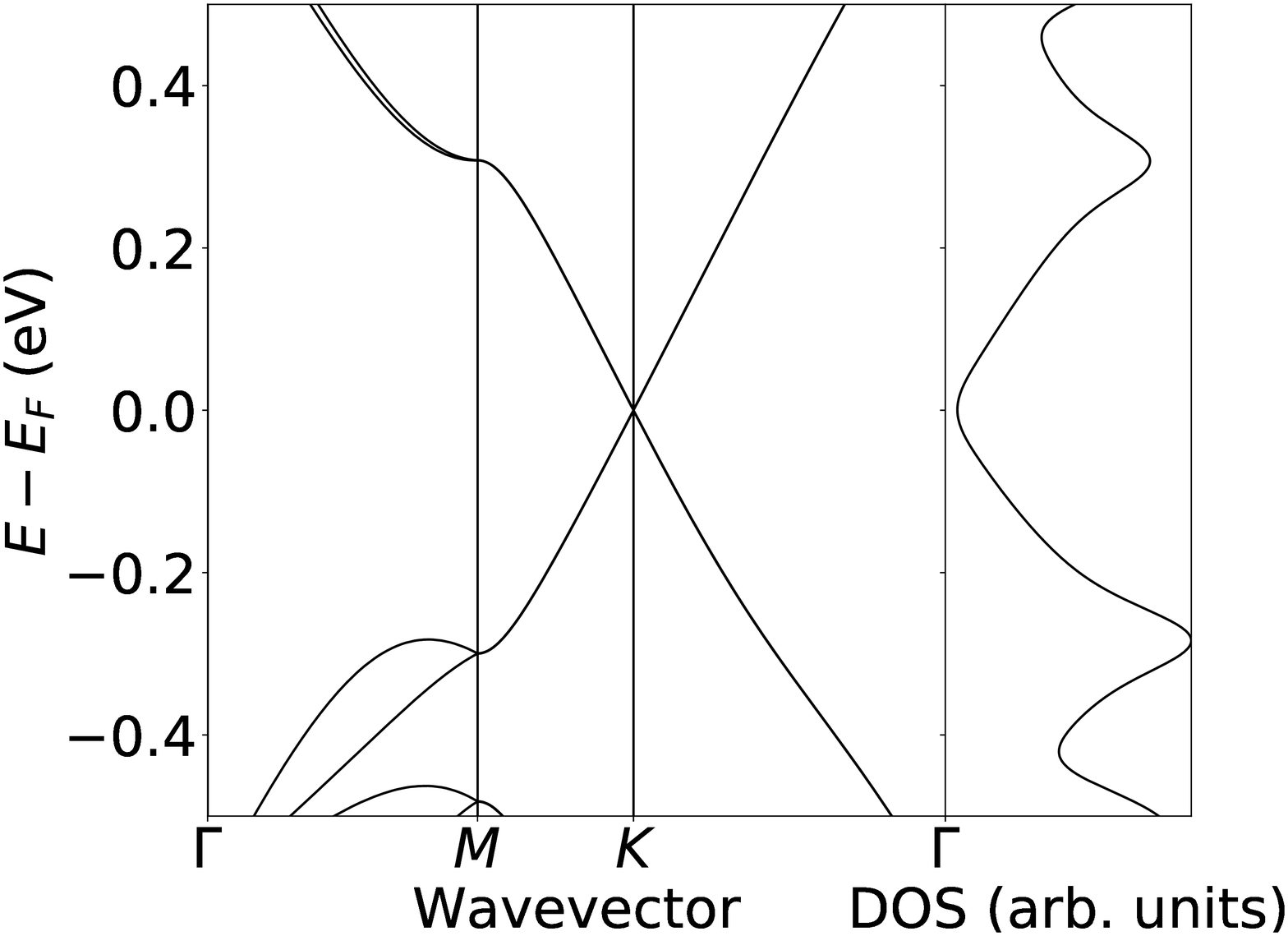}\label{fig:6_7_bands}}%
    \hspace*{-15.0pt}%
    \vspace{-0.5\baselineskip}%
    \subfloat{\subfigimgb[width=0.5\linewidth]{(\textbf{b})}{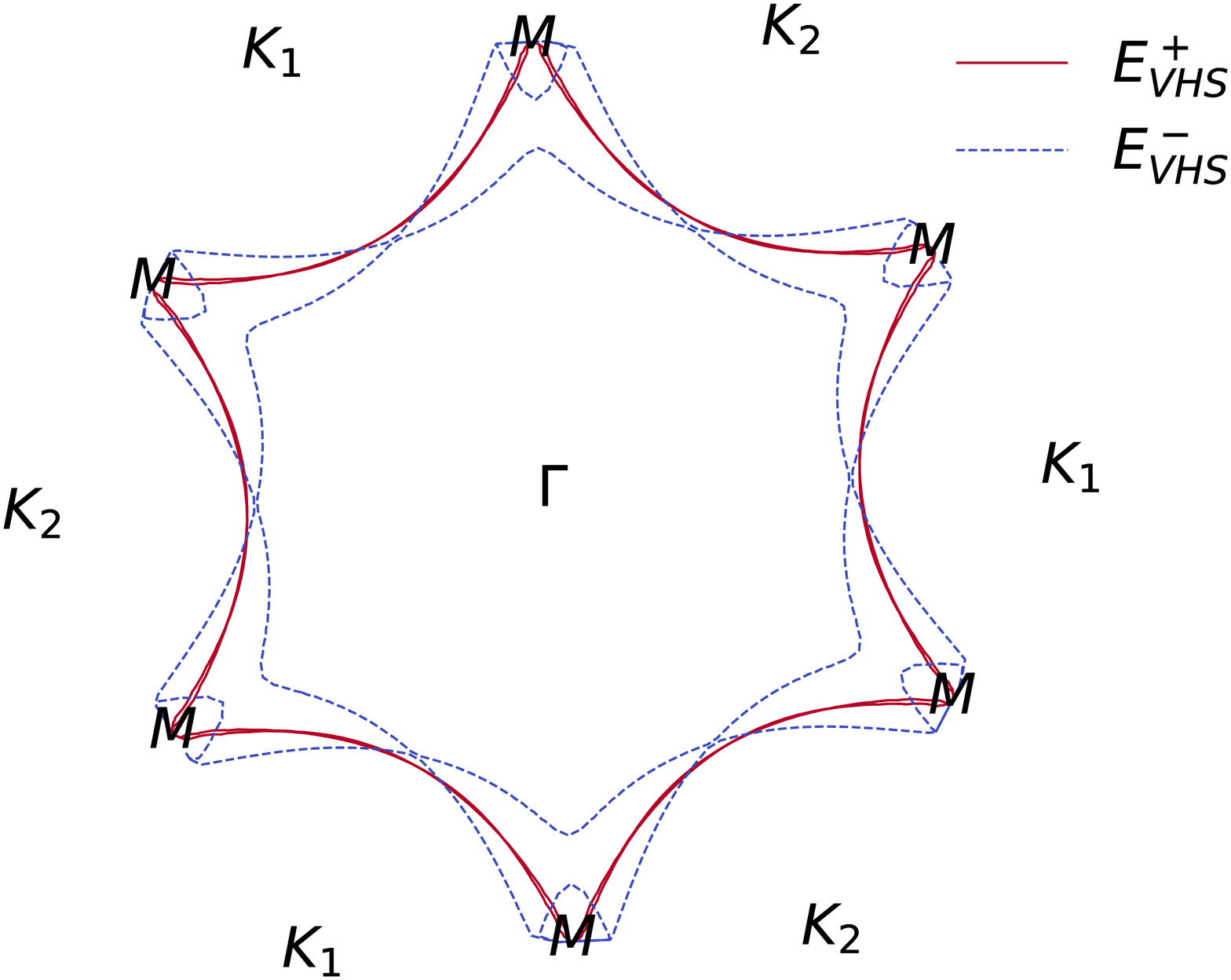}\label{fig:6_7_fs}}%
    \hspace*{-15.0pt}%
    \vspace{-0.5\baselineskip}%
    \\
    \subfloat{\subfigimgb[width=0.5\linewidth]{(\textbf{c})}{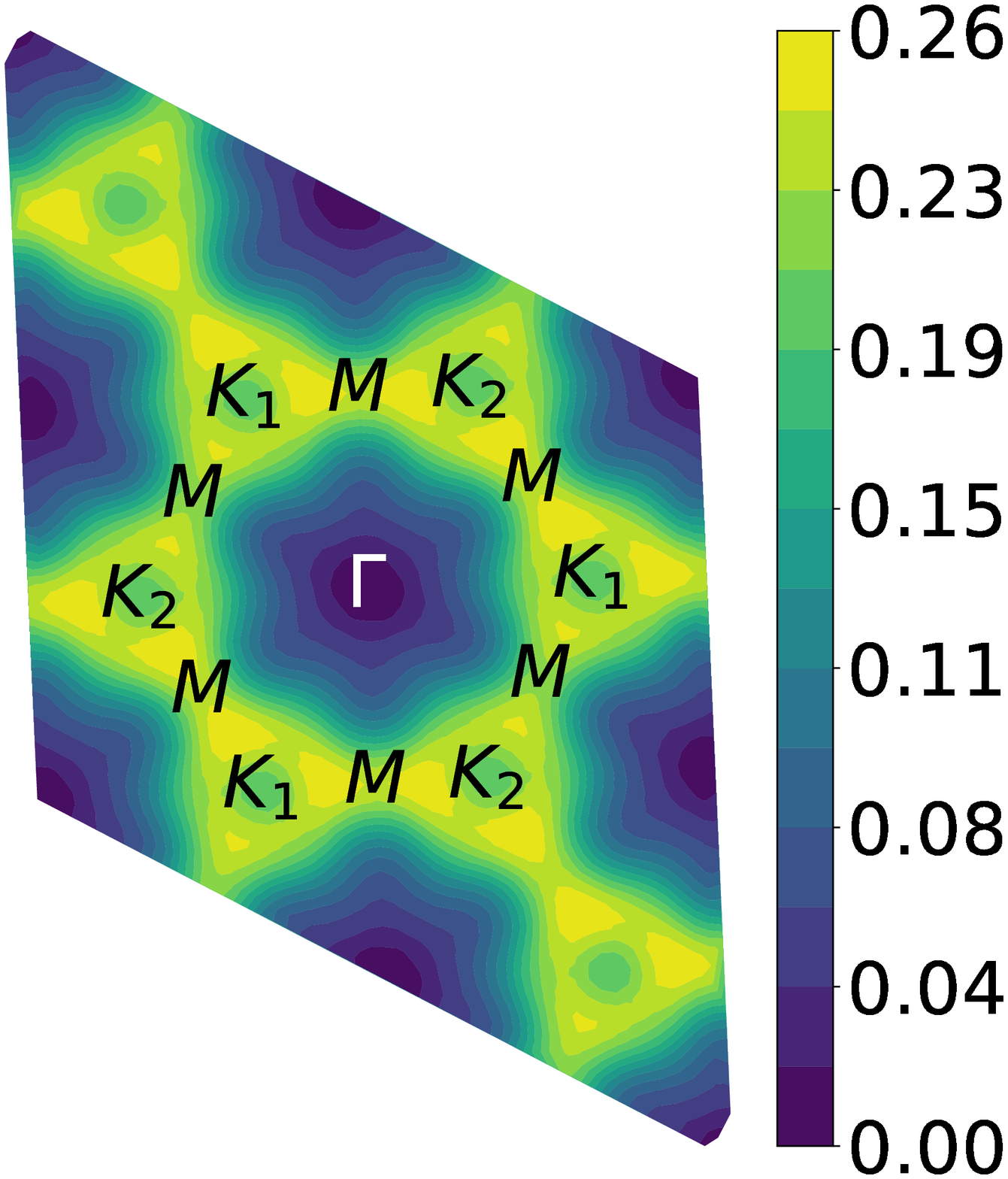}\label{fig:6_7_cm_pos}}%
    \hspace*{-15.0pt}%
    \vspace{-0.5\baselineskip}%
    \subfloat{\subfigimgb[width=0.5\linewidth]{(\textbf{d})}{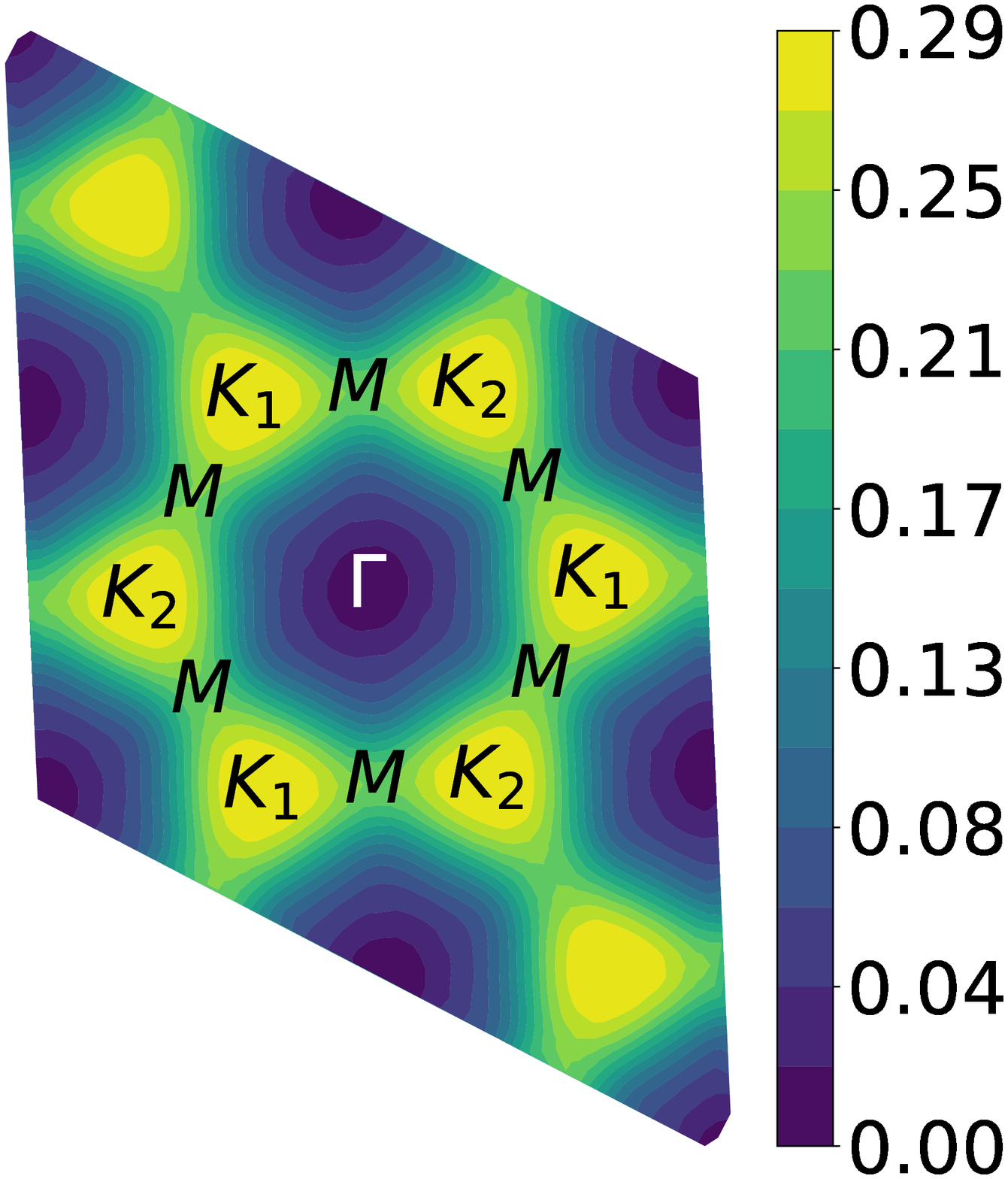}\label{fig:6_7_cm_neg}}%
    \hspace*{-15.0pt}%
    \vspace{-0.5\baselineskip}%
    \\
    \subfloat{\subfigimgb[width=0.5\linewidth]{(\textbf{e})}{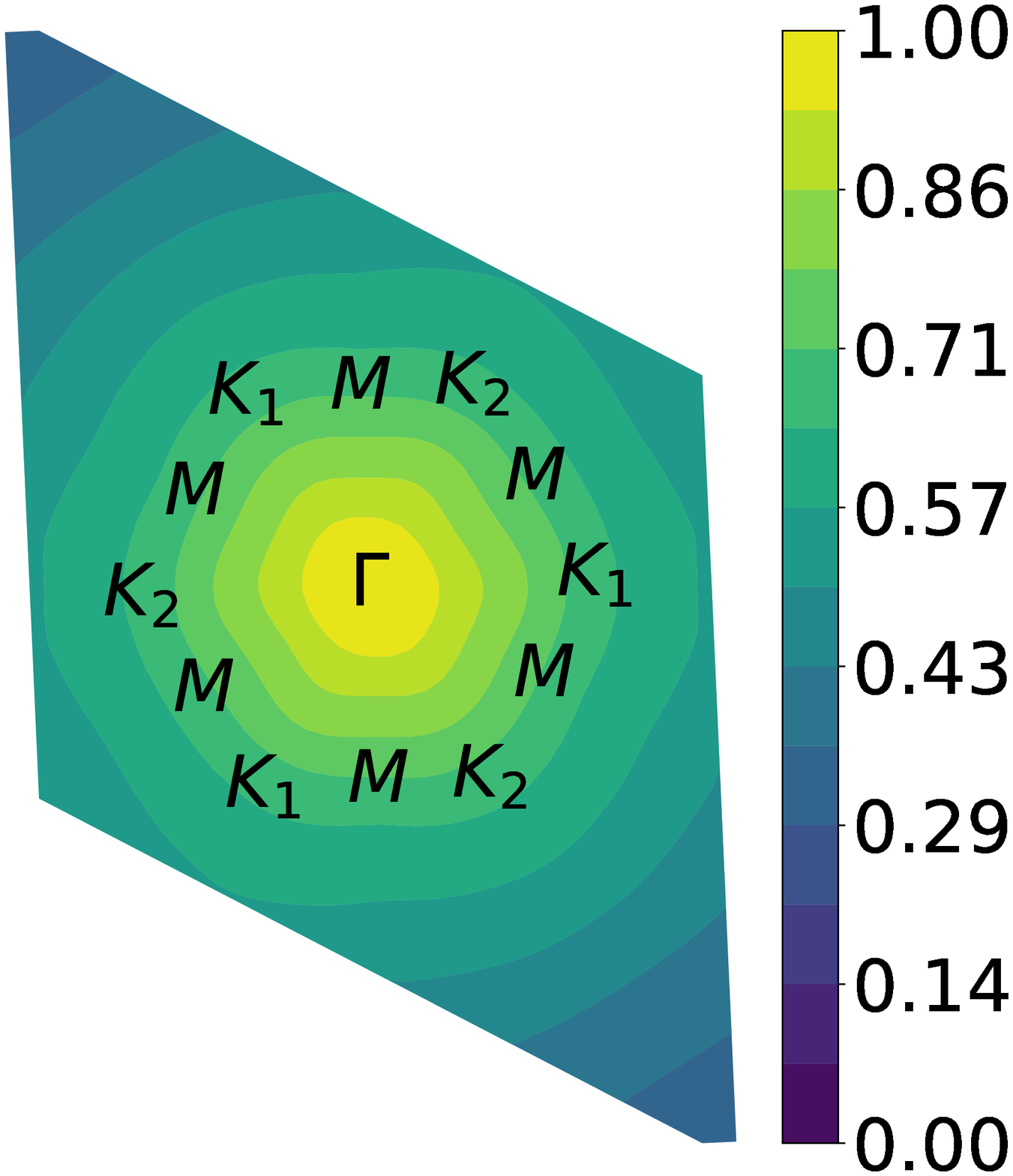}\label{fig:6_7_chi_pos}}%
    \hspace*{-15.0pt}%
    \subfloat{\subfigimgb[width=0.5\linewidth]{(\textbf{f})}{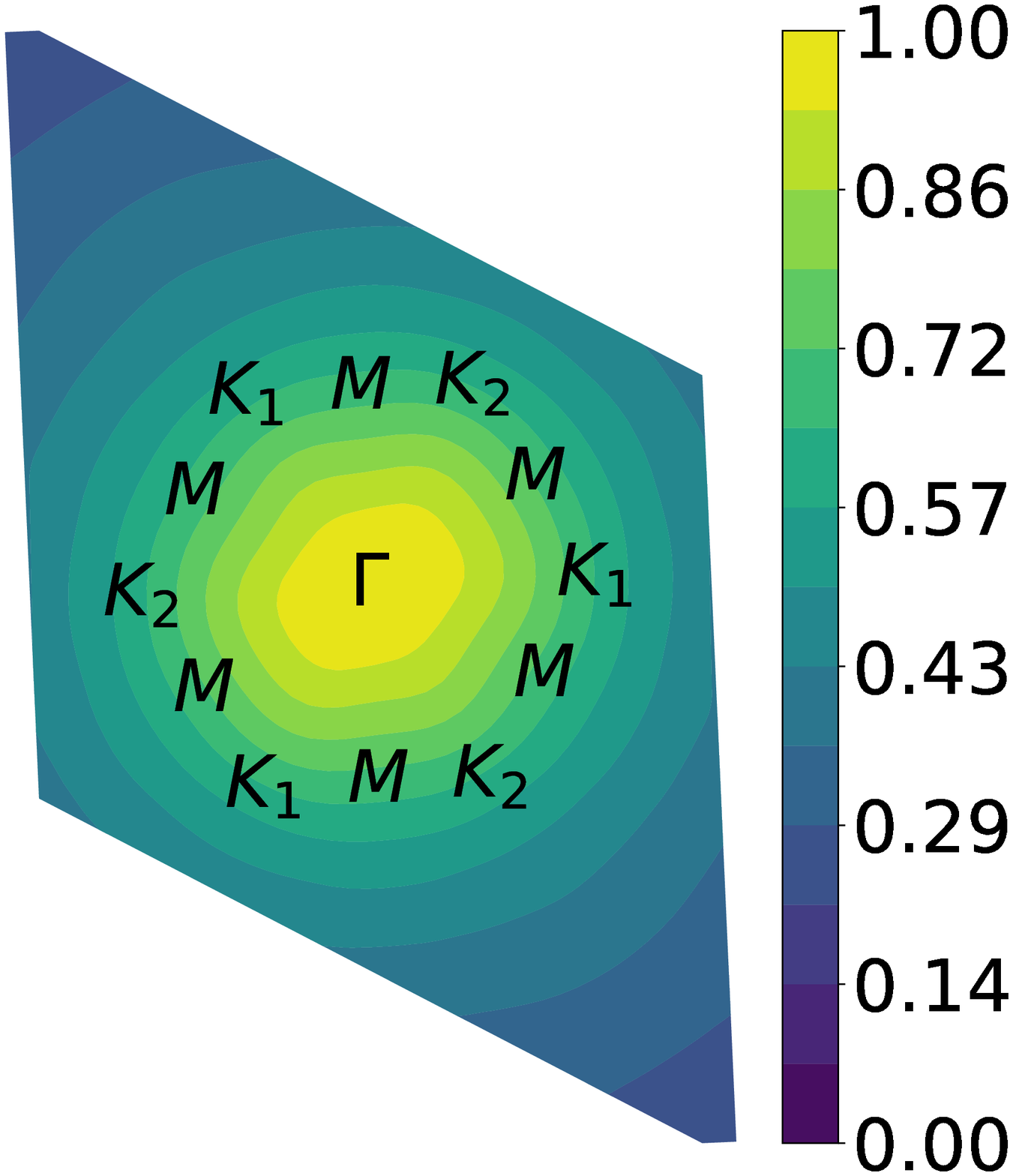}\label{fig:6_7_chi_neg}}%
    \caption{(Color on line) Twisted bilayer graphene response function.  Panels (a) and (b) show the band structure and the Fermi surface for the Fermi level at the two saddle points. Panels (c, d) show the real part of the response function assuming constant matrix elements, where the lowest value was taken as zero. In panels (e, f) the real part of the full response function scaled by $\operatorname{Re}\chi_0(\textbf{q}=0)$ is shown. The Fermi surface and response function were evaluated at (c, e) $E^+_{VHS}$ and (d, f) $E^-_{VHS}$.}
    \label{fig:6_7}
\end{figure}

Twisted bilayer graphene is two stacked layers of graphene, where one graphene layer is rotated relative to the other, forming a Moir\'{e} pattern.
For small rotation angles, known as magic angles, the material undergoes a superconducting transition.\cite{Cao20181,Cao20182}
In the limit of first principles we have performed a calculation for a twist angle of $\theta=5.08^{\circ}$ which has a total of 508 carbon atoms, using the Methfessel-Paxton occupation function,\cite{PhysRevB.40.3616} which allows us to obtain a free energy, very close to the physical energy at $T=0$.
The twist angle is relatively far from the magic angle of $\theta_m\approx1.1^{\circ}$, however the system still exhibits two very large VHS (see Fig. \ref{fig:6_7_bands}), at energies above ($E^+_{VHS}$) and below ($E^-_{VHS}$) the Fermi energy.
%
%we also perform the calculations with hydrostatic pressure, at 1.06 GPa, which should bring our results closer to those of the magic angle.\cite{PhysRevB.99.045423,Yankowitz1059}
%
In this system, we are mainly interested in the interaction between the graphene layers and the VHS, which are meditated by the $\pi$ electrons near the Fermi energy.\cite{PhysRevB.82.121407}
Therefore, we define an energy window [-3, 3] eV in the summation of the bands, so that only the $\pi$ electrons\cite{PhysRevB.77.035427} are included in the calculation of the response function.
We also found that $|\textbf{G}_{\text{cut}}|^{2}=$ 1 Ry is sufficient to achieve a converged result for our purposes.
The plot of the FS, shown in Fig. \ref{fig:6_7_fs}, reveals the nesting vectors are the $M$ points and $K$ points, at $E^+_{VHS}$ and $E^-_{VHS}$, respectively.
Also because the VHS occurs at saddle points there is nesting at the $\Gamma$ point.
Likewise, peaks are observed at those $k$-points in the constant matrix response function evaluated at $E^+_{VHS}$ and $E^-_{VHS}$, as shown in Fig. \ref{fig:6_7_cm_pos} and Fig. \ref{fig:6_7_cm_neg}, respectively.
Although, unlike the hydrogen square lattice, there is no peak at the $\Gamma$ point.
However, when the matrix elements are included, as shown in Fig. \ref{fig:6_7_chi_pos} and Fig. \ref{fig:6_7_chi_neg}, there is only a plateau centered at the $\Gamma$ point that smoothly decays with increasing \textit{q}-vector, destroying any resemblance with the constant matrix response function.
This result heavily resembles the response function for the two-dimensional free electron gas,\cite{1111.5337} which is unsurprising considering that free-standing graphene has a very large in-plane electron mobility.\cite{BOLOTIN2008351}
We also observe that the DOS is larger at $E^-_{VHS}$ than $E^+_{VHS}$, as well as, the response function decays faster at $E^-_{VHS}$ than $E^+_{VHS}$.
Evidently, the large DOS at the VHS is responsible for the rapid decay of the matrix elements that wash away the structure of the constant matrix response function.

%-----------------------------------------

%-----------------------------------------

\subsubsection{Vanadium diselenide monolayer}
\begin{figure} 	
    \centering
    \subfloat{\subfigimgb[width=0.5\linewidth]{(\textbf{a})}{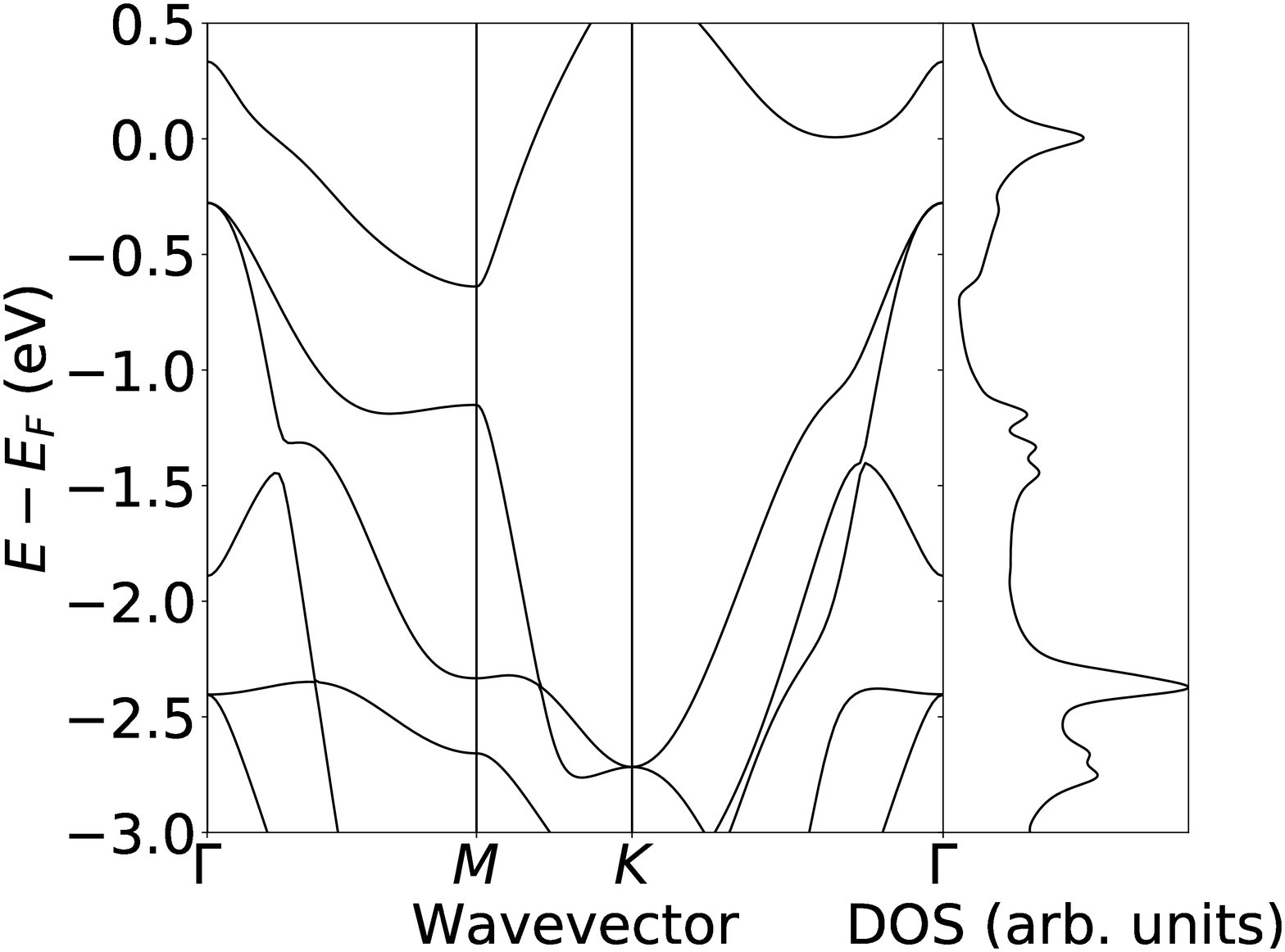}\label{fig:vse2_bands}}%
    \hspace*{-15.0pt}%
    \vspace{-0.5\baselineskip}%
    \subfloat{\subfigimgb[width=0.5\linewidth]{(\textbf{b})}{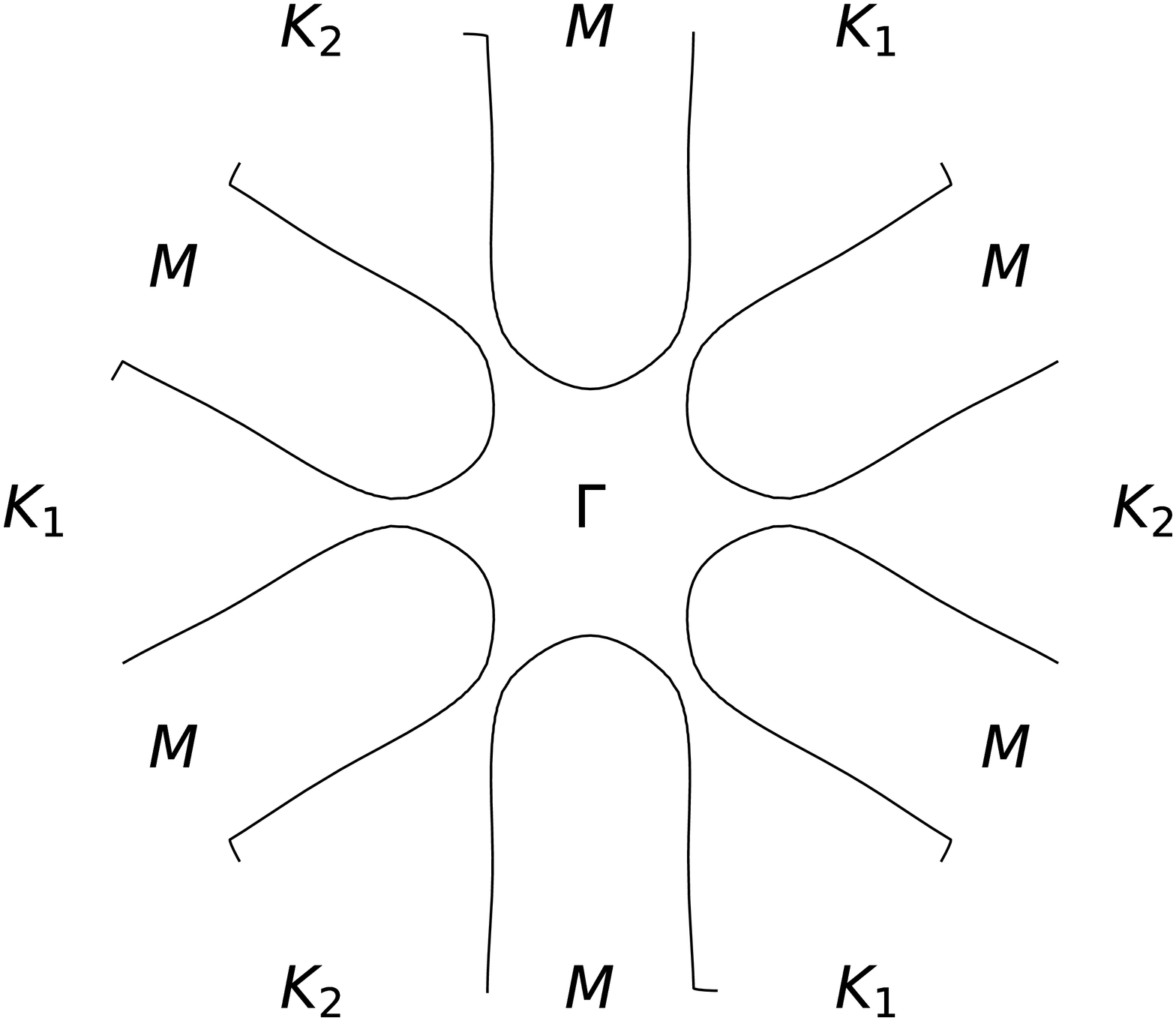}\label{fig:vse2_fs}}%
    \hspace*{-15.0pt}%
    \vspace{-0.5\baselineskip}%
    \\
    \subfloat{\subfigimgb[width=0.5\linewidth]{(\textbf{c})}{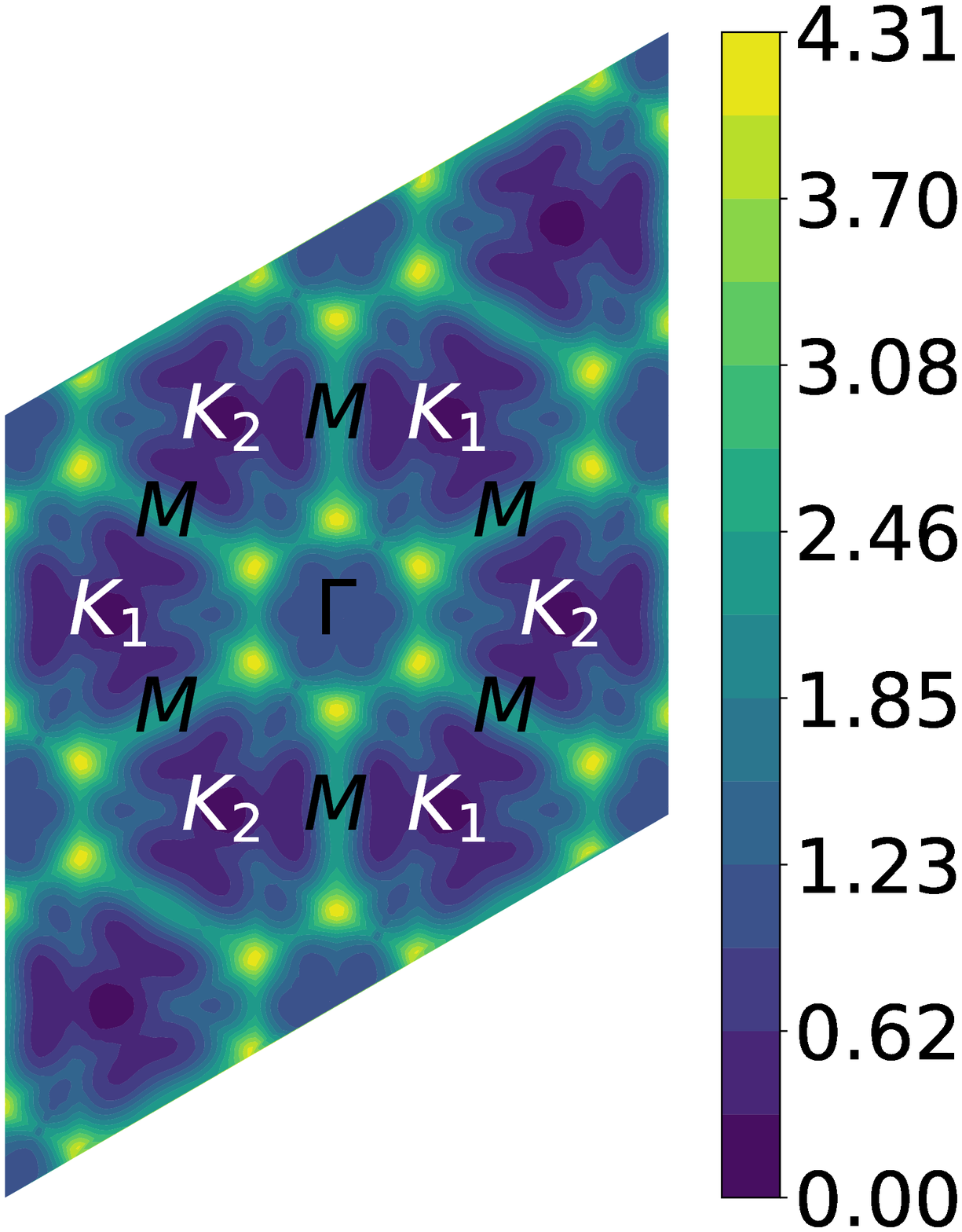}\label{fig:vse2_cm}}%
    \hspace*{-15.0pt}%
    \subfloat{\subfigimgb[width=0.5\linewidth]{(\textbf{d})}{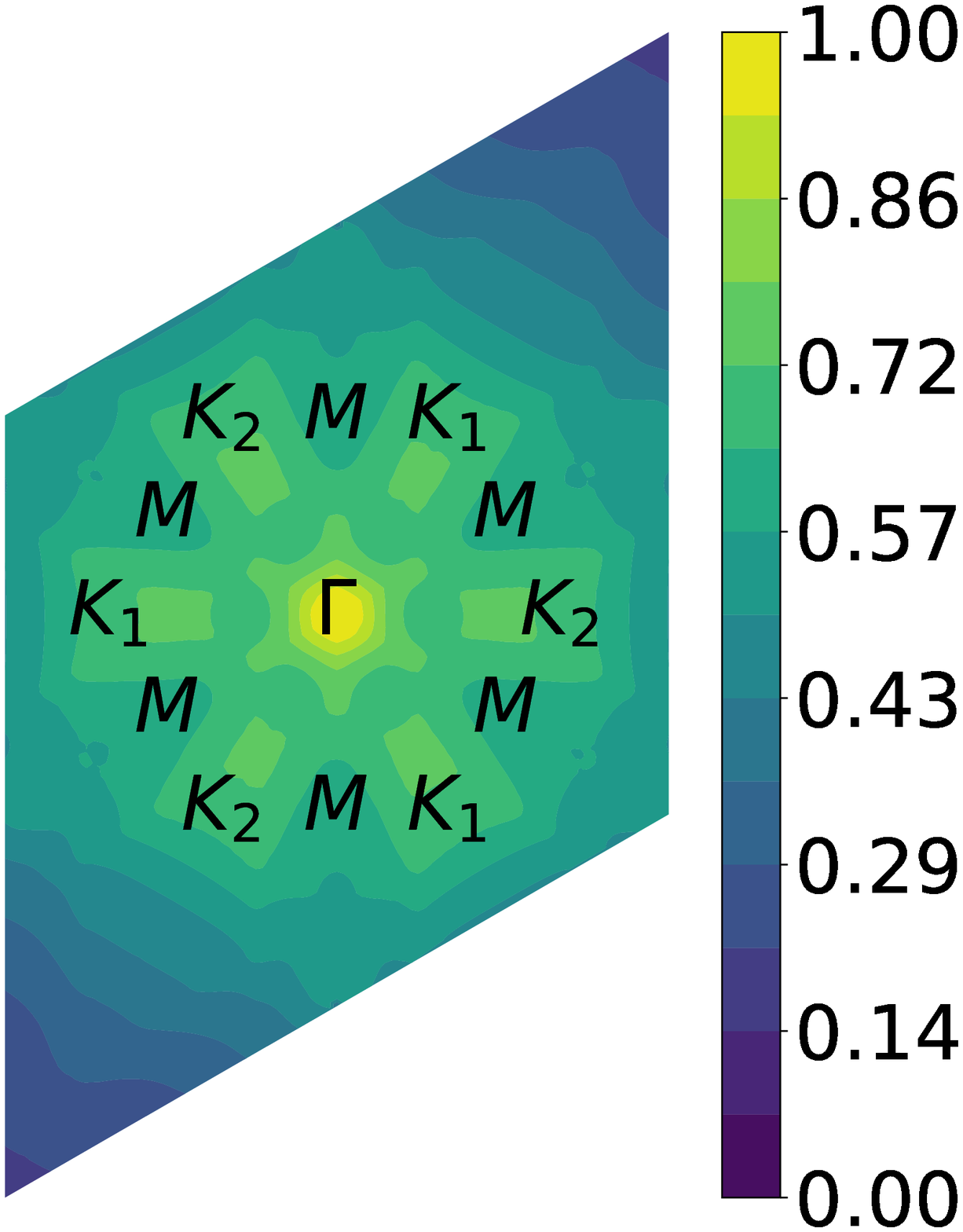}\label{fig:vse2_chi}}%
    \caption{(Color on line) Response function of a VSe\textsubscript{2} monolayer. (a) band structure, (b) the Fermi surface, (c) the real part of response function assuming constant matrix elements, where the lowest value was taken as zero, and (d) the real part of the full response function scaled by $\operatorname{Re}\chi_0(\textbf{q}=0)$. The Fermi surface and response function were evaluated at $\mu=0.0$ eV.}
    \label{fig:vse2}
\end{figure}

The VSe\textsubscript{2} monolayer is a two-dimensional transition-metal dichalcogenide where the charge density wave has been experimentally observed\cite{Feng2018} and leads to the suppression of the ferromagnetic order.\cite{Coelho2019}
The non-magnetic calculation was performed for the free-standing monolayer in the experimentally synthesized 1\textit{T} phase, at the calculated equilibrium lattice constant of 3.39 \AA{}.
The system exhibits a VHS at the Fermi energy, shown in Fig. \ref{fig:vse2_bands} and the nesting vectors of the FS at the VHS are approximately half the distance to the $M$ points, shown in Fig. \ref{fig:vse2_fs}.
We found very sharp peaks in the constant matrix response function, as shown in Fig. \ref{fig:vse2_cm}, however when matrix elements are included, as shown in Fig. \ref{fig:vse2_chi}, there is a large peak at the $\Gamma$ point and smaller satellite peaks, which do not correspond to any peaks in the constant matrix response function.
The largest peak is located at the $\Gamma$ point, rather than the $q$-vectors observed experimentally, suggests that there is an instability associated with the large DOS.\cite{davidpines1989}
The Stoner instability is a natural consequence of this,\cite{10.1098/rspa.1938.0066} and is supported by the fact that the theoretical ground state of this system is ferromagnetic.\cite{Feng2018,Coelho2019}
The existence of the satellite peaks could support the reconstruction of the lattice to accommodate a charge density wave (see Ref. \citenum{Coelho2019}), however this investigation is outside the scope of this work.
Instead, we note that this example highlights that very sharp peaks in the constant matrix response function, even sharper than in the one-dimensional hydrogen chain, are not guaranteed to be preserved in the full calculation.

%-----------------------------------------

			    \label{sec:real}
	\section{Conclusion}
	        We have explored the behavior of the non-interacting response function for both model systems and real materials, all within the framework of DFT. We focused our research on the importance of matrix elements between occupied and empty states since usually, in the literature, they are considered to be constant  like in the free electron case.
The results, for the model systems of the hydrogen chain and two-dimensional square lattice, which were in agreement with model tight-binding calculations, highlighted that even for simple systems the inclusion of matrix elements is essential to map the true instabilities of the material.
The real examples of tBLG and monolayer VSe\textsubscript{2} have shown that neither nesting vectors from the FS nor large peaks in the constant matrix response function are sufficient to predict the peaks in the response function when matrix elements are included.
In addition, we have also elucidated how to perform computationally expensive calculations or isolate orbitals of interest by introducing an energy window for the summation of the bands.

We have observed that a peak in the response function for two-dimensional materials is elusive when $\textbf{q}\neq0$.
In such cases, perfect nesting is not guaranteed, hence the energy loss due to the distortion of the electronic density is not large enough to overcome the energy gain of the lattice distortions.
The interplay between the electronic and lattice distortions warrants further investigation, however it is clear that the interaction is mediated through the matrix elements.
There is still a rich area of physics to explore by generalizing the non-interacting response function to account for spin-polarization and local-field effects.
We hope that future works will seriously consider the calculation of matrix elements when making predictions about the instabilities from the FS or constant matrix response function.
    \section*{Acknowledgements}
            This work was supported by the Spanish Ministry of Science and Innovation through grants
FIS2015-64886-C5-5-P,   MAT2017-83553-P and PGC2018-096955-B-C42.

\FloatBarrier
\bibliography{main}
\end{document}